# Magnetoelectric coupling in the paramagnetic state of a metal-organic framework

W. Wang,[1,2] L.-Q. Yan,[1] J.-Z. Cong,[1] Y.-L. Zhao,[1] F. Wang,[1] S.-P. Shen,[1] T. Zou,[1] D. Zhang,[2] S.-G. Wang,[1] X.-F. Han[1] & Y. Sun[1,*]

[1]State Key Laboratory of Magnetism and Beijing National Laboratory for Condensed Matter Physics, Institute of Physics, Chinese Academy of Sciences, Beijing 100190, P. R. of China

[2]College of Materials and Chemistry, China University of Geosciences, Wuhan, Hubei 430074, P. R. China

*Correspondence and requests for materials should be addressed to Y.S. (email: youngsun@iphy.ac.cn).

Abstract

Although the magnetoelectric effects - the mutual control of electric polarization by magnetic fields and magnetism by electric fields, have been intensively studied in a large number of inorganic compounds and heterostructures, they have been rarely observed in organic materials. Here we demonstrate magnetoelectric coupling in a metal-organic framework $[(CH_3)_2NH_2]Mn(HCOO)_3$ which exhibits an order-disorder type of ferroelectricity below 185 K. The magnetic susceptibility starts to deviate from the Curie-Weiss law at the paraelectric-ferroelectric transition temperature, suggesting an enhancement of short-range magnetic correlation in the ferroelectric state. Electron spin resonance study further confirms that the magnetic state indeed changes following the ferroelectric phase transition. Inversely, the ferroelectric polarization can be improved by applying high magnetic fields. We interpret the magnetoelectric coupling in the paramagnetic state in the metal-organic framework as a consequence of the magnetoelastic effect that modifies both the superexchange interaction and the hydrogen bonding.

Recent years have seen an intense effort to discover new classes of single-phase multiferroics with considerable magnetoelectric (ME) effects[1-4]. While most work are currently focusing on inorganic materials, organic systems such as molecular magnets[5] and organic ferroelectrics[6] could open up new routes to multiferroics and ME effects. For instance, magnetic control of ferroelectricity has been observed at low temperature in a one-dimensional organic magnet[7]. In addition to pure inorganic and organic materials, multiferroics may be achieved in hybrid metal-organic frameworks (MOFs) that combine magnetism of metal ions with organic ferroelectricity[8-10]. A good example is the MOFs with a general formula $[(CH_3)_2NH_2]M(HCOO)_3$, where M=divalent metal ion[11]. These MOFs experience a ferroelectric (or antiferroelectric) phase transition in the temperature range of 160 - 185 K, depending on the specific M, due to hydrogen bond ordering[12-14]. As they also exhibit a canted antiferromagnetic ordering at low temperatures[15,16] ($T_c$=8 - 36 K) with M=$Mn^{2+}$, $Co^{2+}$, $Ni^{2+}$ and $Fe^{2+}$, the $[(CH_3)_2NH_2]M(HCOO)_3$ family represents a new class of multiferroics[11]. However, these MOF materials belong to the type-I multiferroics[17] where the magnetic and ferroelectric orders have different origins. Thus, one may not expect a pronounced ME effect in the multiferroic phase. Nevertheless, the ME effects are not restricted to the multiferroic state; it may exist whatever the nature of magnetic order parameters, and can for example occur in paramagnetic ferroelectrics[2,18,19].

In this work, we demonstrate the ME coupling in the paramagnetic state near the ferroelectric phase transition in the MOF materials, $[(CH_3)_2NH_2]Mn(HCOO)_3$. The magnetic susceptibility and electron spin resonance (ESR) spectra clearly evidence that the magnetic state changes following the ferroelectric phase transition. Inversely, the ferroelectric state can be modified by applying a high magnetic field. Such a ME coupling in the paramagnetic state is probably related to the magnetoelastic effect that modifies both the hydrogen bonding and the superexchange interaction between $Mn^{2+}$ ions.

**Results**

**Characterization of $[(CH_3)_2NH_2]Mn(HCOO)_3$ single crystal.** As shown in Fig. 1a, this MOF material has a $ABX_3$ perovskite architecture, in which the metal cations ($B=Mn^{2+}$) linked by formate groups ($X=HCOO-$) form the $BX_3$ skeleton, and the dimethylammonium (DMA) cations occupy the cavities ($A=[(CH_3)_2NH_2]^+$). The amine hydrogen atoms make hydrogen bonds with the oxygen atoms of the metal formate framework. Because of this hydrogen bonding, the nitrogen of the ammonium cation is disordered over three equal positions at room temperature.

The prepared single crystals of $[(CH_3)_2NH_2]Mn(HCOO)_3$ were characterized by powder X-ray diffraction (XRD), single crystal XRD, magnetic and dielectric property measurements. The powder XRD pattern at room temperature (Supplementary Fig. S1) confirmed the structure and phase purity of the obtained samples. The single crystal XRD pattern (Supplementary Fig. S1) suggests that the crystal is naturally grown layer by layer along the [012] direction. Fig. 1b shows the magnetization as a function of temperature measured in 0.1 T. A magnetic phase transition is clearly seen at 8.5 K. Moreover, a magnetic anisotropy is observed in the single crystal sample. The magnetization along the [012] direction drops slowly with decreasing temperature while the magnetization perpendicular to [012] exhibits a jump at 8.5 K and remains high down to 2 K. The angular dependence of magnetization at 2 K (see Supplementary Fig. S2) also suggests that the magnetic easy axis lies in the (012) plane. However, the M-H loops measured between ±7 T at 2 K shown in the inset of Fig. 1b are almost identical for both directions, except a small difference in the low field ranges. The linear shape of the M-H loops and the small hysteresis at low fields indicate a canted antiferromagnetic structure. Moreover, the ESR spectra at low temperature (T<8.5 K) exhibit two separate resonance lines (Fig. 3a), which is also a characteristic of antiferromagnetism.

Fig. 2a represents the dielectric constant along [012] direction as a function of temperature. With increasing temperature, the dielectric constant increases slowly at low temperature and exhibits a sudden, big jump at 185 K. Meanwhile, the dielectric loss also shows a sharp peak at the same temperature (inset of Fig. 2a). We note that the rise of dielectric loss near 185 K is not due to leakage current because the MOF sample is highly insulating, with a resistivity ~ $2\times10^{10}$ ohm.cm at room temperature and over $10^{12}$ ohm.cm below 200 K. These dielectric anomalies indicate a first-order like phase transition at 185 K. Initially, this transition was initially assumed to be the paraelectric-antiferroelectric type[12]. More recently, it has been proposed that this transition is associated with the appearance of improper ferroeletricity[10]. The structural, magnetic, and dielectric properties are all consistent with literature, which suggests the high quality of our single-crystal samples. Since the system enters into a multiferroic state, *i.e.*, the coexistence of ferroelectricity and magnetic ordering, below 8.5 K, one may wonder if there is ME effects in the multiferroic state. We have testified it by measuring the dielectric response along [012] direction at low temperature across the magnetic transition. The dielectric constant did not show a clear anomaly around the magnetic ordering temperature within the measurement resolution of our LCR meter. These preliminary results seem negative for the ME effect. However, it is still inconclusive at this stage whether there is ME coupling in the multiferroic state and further investigation is required.

**Correlation between magnetism and ferroelectricity in the paramagnetic state.** In previous studies it has been generally assumed that the MOF system is in a simple paramagnetic state at high temperature (T > 8.5 K). However, when we check carefully the high temperature magnetic susceptibility of single crystal samples, an apparent anomaly is observed around the ferroelectric transition temperature, $T_c \sim 185$ K. As shown in Fig. 2b, the inverse magnetic susceptibility, $\chi^{-1}$, follows well the Curie-Weiss law, $\chi=C/(T-\theta)$, at high temperature. Just at the ferroelectric $T_c$, it starts to deviate from the Curie-Weiss law and the discrepancy becomes notable at low temperature. As the $\chi^{-1}$ is lower than the Curie-Weiss line, it implies that there are short-range magnetic correlations that give a higher Curie constant, $C$. In other words, the short-range magnetic correlation is apparently enhanced in the ferroelectric state. Thus, the anomaly in magnetic susceptibility at the ferroelectric Curie temperature, $T_c$, suggests a possible ME coupling.

To further investigate the magnetic states below and above $T_c$, we employed ESR technique that is very sensitive to the change of magnetic states. Fig. 3a plots the ESR spectra at several representative temperatures. A well-defined single paramagnetic resonance line is observed above 8.5 K. The temperature dependence of the ESR intensity provides in-depth information on the nature and variation of the magnetic state. The increase in intensity with decreasing temperature usually exhibits the $T^{-1}$ Curie-like behavior expected for a single-ion excitation ($Mn^{2+}$), whereas the $(T-\theta)^{-1}$ Curie-Weiss law is commonly observed for exchange coupled ions[20,21]. As shown in Fig. 3b, the ESR intensity vs inverse temperature of the MOF single crystal follows the Curie-Weiss law in the low temperature regime ($T < 185$ K). This temperature dependence of ESR intensity is consistent with what expected for short-range exchange coupled $Mn^{2+}$ ions. Moreover, the ESR intensity exhibits a sudden drop near $T_c \sim 185$ K. The smaller ESR intensity at elevated temperatures than the low-T Curie-Weiss line implies that the short-range magnetic correlation is suppressed in the paraelectric phase. These apparent anomalies in the ESR spectra further evidence that the magnetic state indeed changes after the paraelectric-ferroelectric transition. Therefore, a ME coupling does exist in the paramagnetic state.

**Magnetic field control of electric polarization.** The cross coupling between the ferroelectric and magnetic states can be further proved by the direct ME effect. Fig. 4a and 4b show the temperature dependence of the pyroelectric current and the integrated electric polarization along [012], respectively. The giant, sharp peak in the pyroelectric current around 185 K reveals a clear ferroelectric phase transition. In order to detect the influence of magnetic fields on intrinsic pyroelectric response, we used the field cooling mode in which the magnetic field is applied with poling E field along [012] at 240 K and kept constant during the cooling and warming process. In this way, the trapping and release of charges across the ferroelectric phase transition happens in the same resistive state, and thus eliminate the possible contribution from the magnetoresistance effect.

As shown in the insets of Fig. 4a, the pyroelectric current is increased by applied 7 T magnetic field, especially in the vicinity of $T_c$, and the peak temperature shifts slightly to high temperature. As a result, the electric polarization in the FE state is enhanced by applying high magnetic fields (Fig. 4b). This direct ME effect (magnetic field control of electric polarization) is consistent with what expected from the magnetic susceptibility and ESR spectra. Since the ferroelectric state favors short-range magnetic interaction, a high magnetic field which strengths the magnetic interaction would promote the ferroelectric state in return. We also tried to measure the *P-E* hysteresis loop of the MOF sample using a positive up negative down (PUND) method[22,23] (see supplementary Fig. S3). Though the results obtained at 500 Hz with applied E field up to 30 kV/cm do not give a clear *P-E* hysteresis loop, probably due to the high coercive electric field required by the MOF and the slow motion of ferroelectric domains, these measurements do confirm that there is an intrinsic remanent polarization.

**Discussion** All above experimental results support the existence of certain coupling between magnetism and ferroeletricity in the paramagnetic state of the MOF. The ME coupling in the paramagnetic state is unusual and there have been only very limited reports so far. The first example was reported by Hou and Bloembergen in 1964 in a piezoelectric paramagnetic crystal of $NiSO_4 \cdot 6H_2O$ (Ref. 18). They termed this kind of paramagnetic magnetoelectric coupling as paramagnetoelectric (PME) effects and proposed that it may appear in other piezoelectric paramagnetic crystal where the PME effect at low temperaure is dominated by the variation of crystal field splitting D with electric field. In early 1990s, magnetic field influence on the electric polarization were observed in the ferroelectric-ferroelastic paramagnetic phase of rare-earth molybdates such as $Tb_2(MoO_4)_3$ and $Gd_2(MoO_4)_3$ (Ref. 19). The authors ascribed this ME effect to the magnetostriction associated with $Tb^{3+}$ and $Gd^{3+}$ ions along with the piezoelectric effect of the ferroelectric substance. Similarly, the ME effect in the MOF could be also related to the magntostriction (magnetoelastic) effect.

As previous studies have clarified, the ferroelectricity in the present MOF is associated with hydrogen bond ordering[12-14]. In the high temperature paraelectric phase, the DMA cations in the cavities are dynamically disordered with nitrogen distributed over three equivalent positions, because the hydrogen bonding between the hydrogen atoms of the $NH_2$ group and oxygen atoms from the formate framework is disordered. Below 185 K, the ordering of nitrogen atoms due to the hydrogen bond ordering of the DMA cations leads to a lowering in symmetry. As a consequence, a structural transition from the rhombohedral to monoclinic symmetry and Cc space group (Cc belongs to one of the 10 polar point groups required for ferroelectricity) is accompanying with the hydrogen bond ordering and the ferroelectric phase transition. The magnetic susceptibility and ESR data suggest that the monoclinic crystalline structure favors the short-range superexchange interaction between $Mn^{2+}$ ions than the rhombohedral structure, possibly due to the modification in the Mn-Mn distance and angle. The correlation between exchange interaction and lattice structure is known as the magnetostriction or magnetoelastic effect. As a prototype of such an effect, the magnetostriction induced ME coupling has been proposed and observed in a number of inorganic oxides[24,25]. Meanwhile, the magnetoelastic coupling is also observed in some organic materials[26,27]. Especially, a recent study by Thomson *et al.* using resonant ultrasound spectroscopy reveals that there is certain magnetoelastic coupling in the MOF family[28]. It is quite likely here in the MOF that the balance between superexchange and elastic energy leads to local distortion that modifies the hydrogen bond ordering and consequently the ferroelectricity. Compared with the conventional magnetostriction effect in long-range magnetic ordering phase, this local magnetoelastic interaction in the paramagnetic state is much weaker, and thus requires high magnetic fields to amplify it.

Although the ME coupling in the paramagnetic state associated with the ferroelectric phase transition in the present MOF system is not pronounced and requires quite high magnetic fields, it happens at relatively high temperature and the electric polarization is 2-3 orders higher than that in magnetic-order-induced type-II multiferroics[4]. Therefore, further exploration and amplification of the ME coupling by designing various MOFs may bring new routes toward high-temperature sensitive ME effects.

**Methods**

**Sample preparation.** Single crystal samples of $[(CH_3)_2NH_2]Mn(HCOO)_3$ were prepared by solvothermal condition method. A 30 mL DMF solution containing 5 mmol metal chloride salts and 30 mL deionized water was heated in a polyphenyl (PPL)-lined autoclaves for 3 days at 140 °C. The supernatants were transferred into a glass beaker to crystallization at room temperature, and cubic colorless crystals were obtained after 3 days. The crystals were filtered from the mother liquid and washed by ethanol. The X-ray diffraction experiments of powder and single crystals were performed at room temperature using a Rigaku x-ray diffractometer.

**Magnetic, electric, and ESR measurements.** A single crystal showing large faces perpendicular to [012] with a size of $2.20 \times 2.30 \times 0.60$ mm$^3$ was used in the magnetic and electric measurements. The magnetic properties were performed on a superconducting quantum interference device magnetometer (Quantum Design MPMS XL-7). A home-made sample probe was adopted into Quantum Design PPMS to do the dielectric and pyroelectric experiments. The electrodes were made with silver paste painted on the top and bottom plate faces. The dielectric response was measured by using a HIOKI 3532-50 LCR meter at several frequencies using an excitation of 1 V. For the polarization measurements, the sample was cooled down from 210 K to 150 K with a poling electric field of 700 kV/m. After removing the poling electric field and waiting for 20 minutes, the pyroelectric current was measured with an electrometer (Keithley 6517B) at a constant warming rate of 2 K/min. Electric polarization ($P$) value was obtained by integrating $I$ with respect to time. The ESR experiments were carried out with a JEOL JES-FA200 ESR spectrometer at X-band frequencies ($v \approx 9.4$ GHz).

magnetoelastic coupling, and order-disorder processes in multiferroic metal-organic frameworks. *Phys. Rev. B* **86**, 214304 (2012).

**Acknowledgments**

The authors are grateful to Dr. Y. S. Chai and Dr. J. Lu for their help in PUND experiments. This work was supported by the National Key Basic Research Program of China under Grant No. 2011CB921801 and the Natural Science Foundation of China under Grants Nos. 11074293, 51021061, and 11227405.

**Additional Information**

**Author contributions**

Y.S., W.W., and L.Q.Y. conceived and designed the experiments. W.W. and L.Q.Y. prepared the samples. W.W., J.Z.C., Y.L.Z., F.W., and S.P.S. performed the experiments. T.Z., D.Z., S.G.W., and X.F.H. contributed material and analysis. Y.S. and W.W. co-wrote the paper.

**Competing financial interests**

The authors declare no competing financial interests.

**Figure legends**

**Figure 1 Characterization of [(CH$_3$)$_2$NH$_2$]Mn(HCOO)$_3$ single crystal.** **(a)** Schematic crystal structure. **(b)** Low temperature magnetization along two crystal directions. A magnetic transition is observed at 8.5 K. The inset shows the M-H curves at 2 K which suggests a canted antiferromagnetic structure.

**Figure 2 Correlation between magnetism and ferroelectricity.** **(a)** Dielectric constant (ε) along [012] direction as a function of temperature. The big jump of ε and the sharp peak in the dielectric loss (shown in the inset) evidence the paraelectric (PE) to ferroelectric (FE) phase transition at $T_c$ ~ 185 K. **(b)** The inverse magnetic susceptibility as a function of temperature. The blue solid line is the fit to the Curie-Weiss law. Just at $T_c$, the susceptibility starts to deviate from the Curie-Weiss law, indicating an enhancement of short-range magnetic correlation in the FE phase.

**Figure 3 Electron spin resonance spectra.** **(a)** The X-band ESR spectra at selected temperatures. A single paramagnetic resonance line is observed above 8.5 K while two resonance lines are observed below 8.5 K. **(b)** Temperature dependence of the ESR intensity. The blue solid line is the fit to the Curie-Weiss law. The deviation from the Curie-Weiss line at $T_c$ indicate a change in the magnetic state following the PE-FE transition.

**Figure 4 Magnetic field control of electric polarization**. **(a)** Temperature dependence of the pyroelectric current in zero and 7 T magnetic field. The sharp peak corresponds to the PE-FE transition. The insets show the enlarged views below and around $T_c$. The peak is enhanced and slightly shifted to higher temperature by applying a high magnetic field. **(b)** The electric polarization along [012] in 0, 7, and 13 T magnetic fields. The ferroelectricity is improved by applying high magnetic fields. The inset shows the measurement configuration. Both the poling electric field and the magnetic field were applied along [012] during the cooling process.

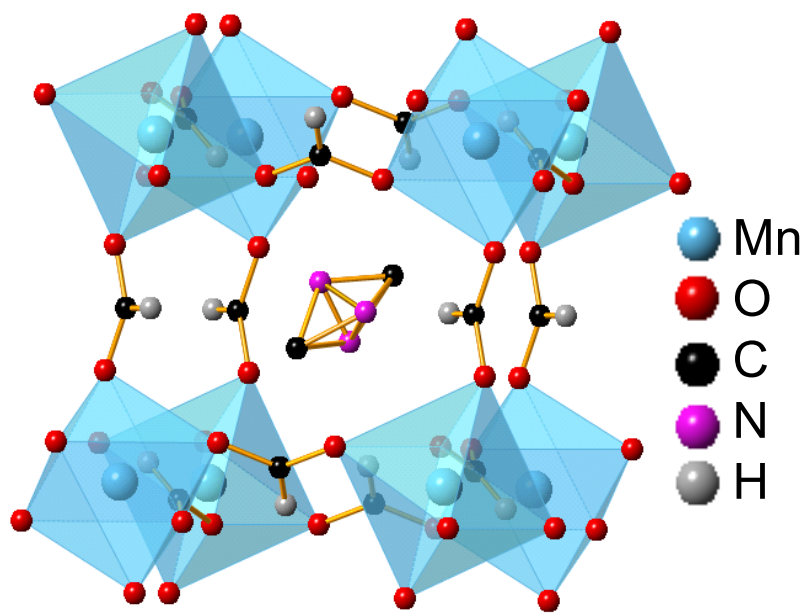
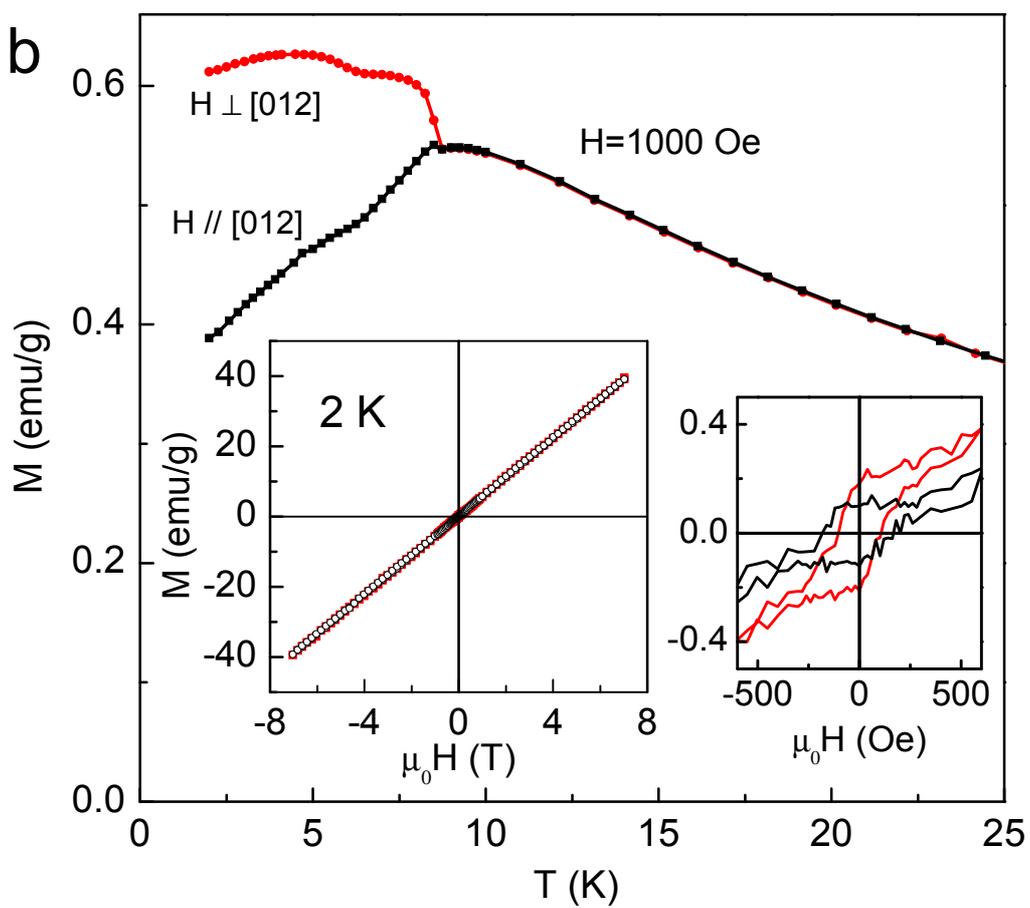

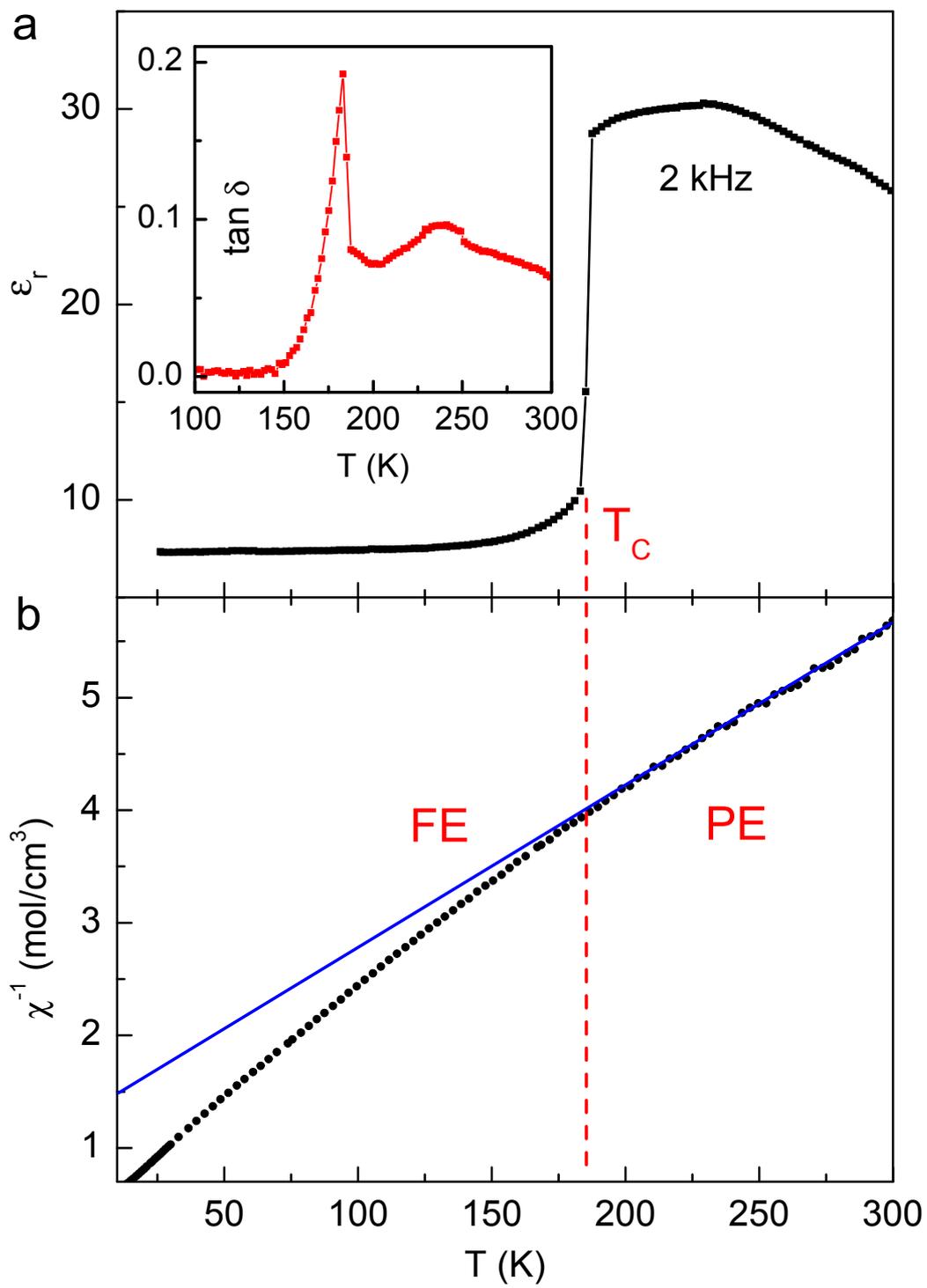

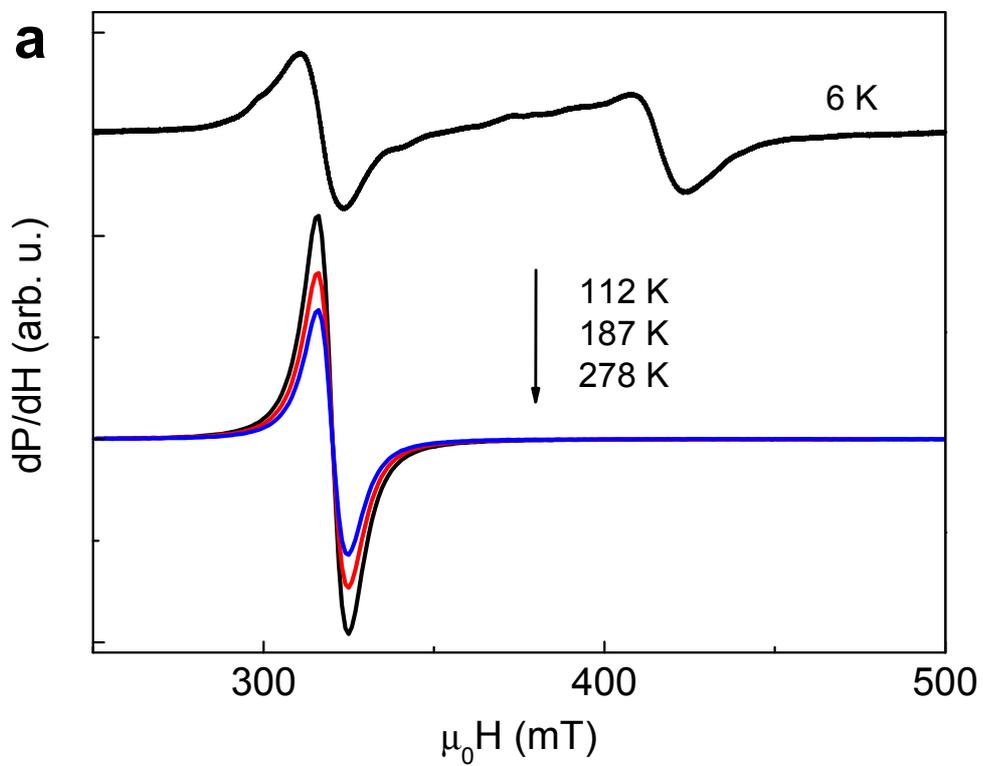
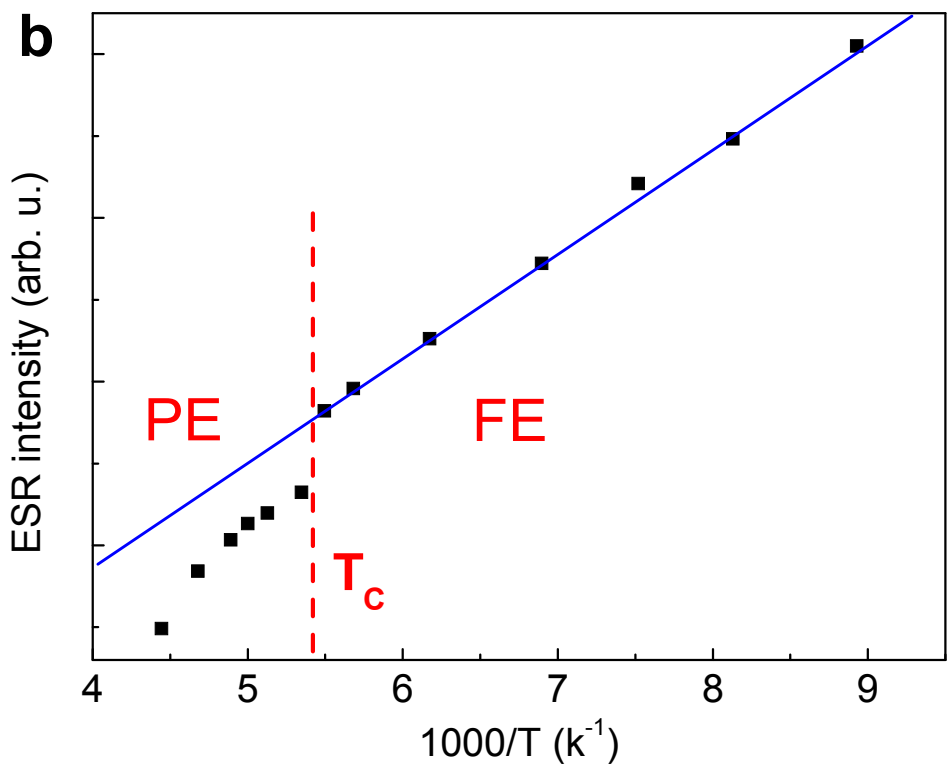

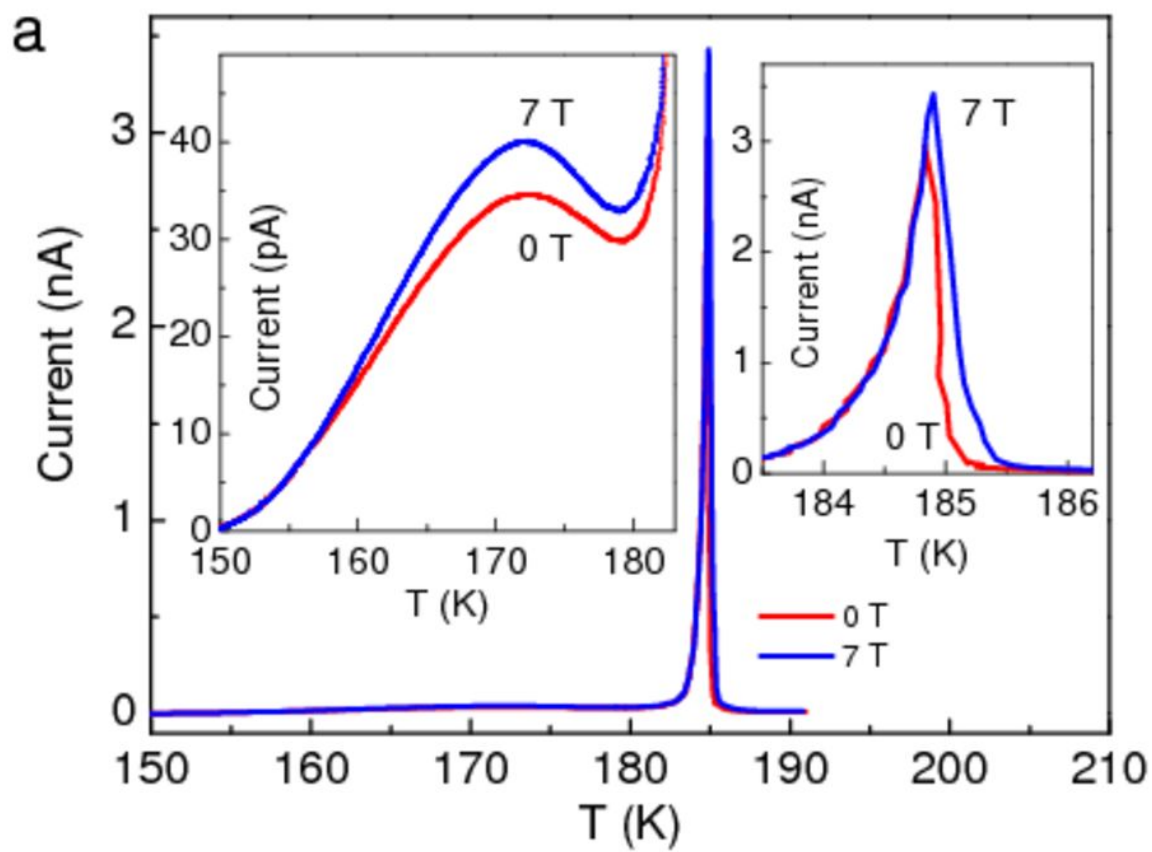

a

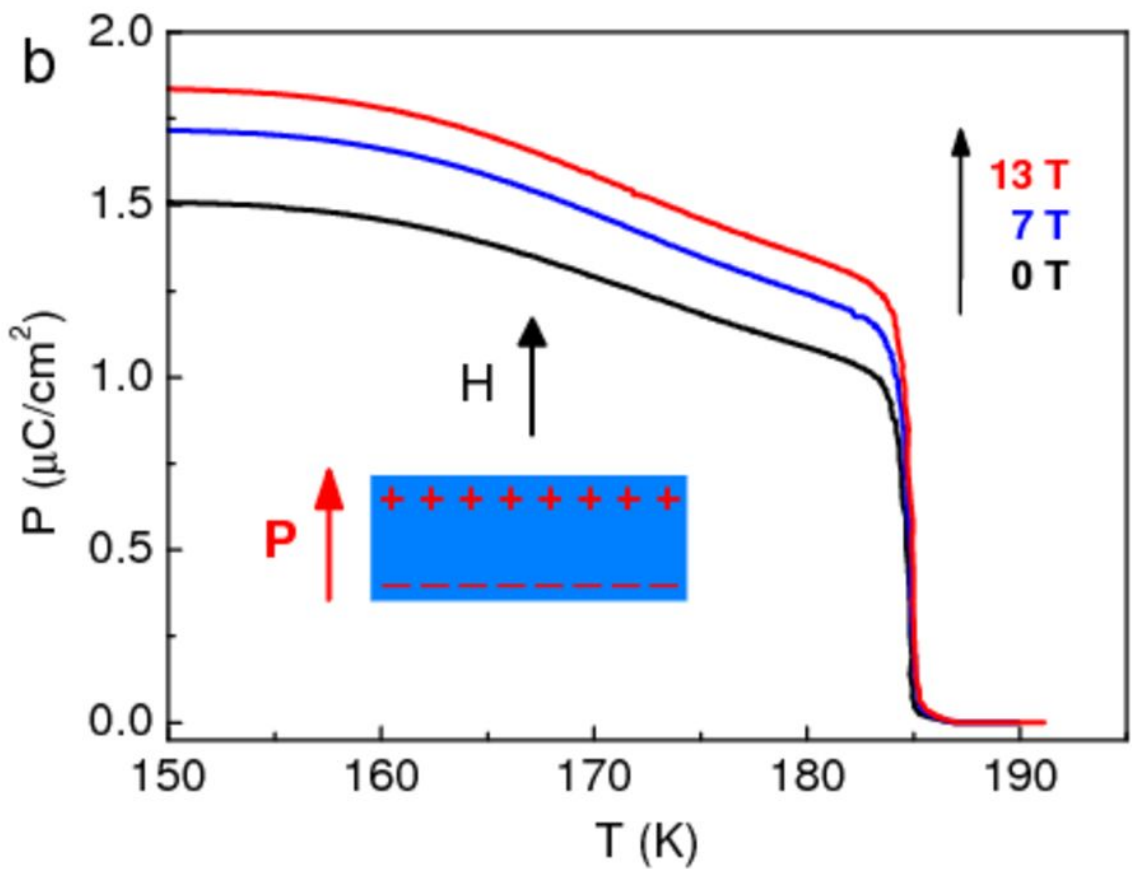

b